\documentclass[twocolumn, pre, showpacs, english, preprintnumbers, amsmath, amssymb, superscriptaddress, aps]{revtex4-2}
\usepackage[subpreambles=true]{standalone}

\usepackage[utf8]{inputenc}
\usepackage{graphicx}
\usepackage{grffile}
\usepackage[usenames,dvipsnames]{xcolor}
\usepackage{amsmath}
\usepackage[normalem]{ulem}
\usepackage[resetlabels, labeled]{multibib}
\usepackage{appendix}
\newcites{S}{References Supplementary Materials}
\definecolor{orange}{rgb}{1,0.5,0}
\definecolor{goodgreen}{rgb}{0.1,0.5,0}
\definecolor{goodred}{rgb}{0.7,0,0}
\usepackage{lineno}
\setpagewiselinenumbers
\usepackage{soul}

\usepackage{tikz}
\usepackage{tocvsec2}
\usepackage{scrextend}
\makeatletter

\usepackage[colorlinks,urlcolor=goodgreen,citecolor=blue,linkcolor=goodred]{hyperref}

\usepackage{physics}

\usepackage{float}
\graphicspath{ {images/} }

\let\oldphi\phi \let\phi\varphi \let\varphi\oldphi
\let\oldepsilon\epsilon \let\epsilon\varepsilon \let\varepsilon\oldepsilon

\makeatother

\usepackage{babel}

\begin{document}

%\title{Manifestation of the Josephson anomalous phase on quasiparticle current in an Andreev interferometer}

\title{Anomalous Andreev interferometer: Study of an anomalous Josephson junction coupled to a normal wire}

\newcommand{\orcid}[1]{\href{https://orcid.org/#1}{\includegraphics[width=8pt]{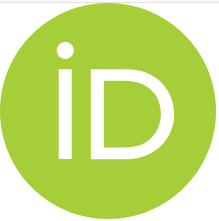}}}

\author{Alberto Hijano\orcid{0000-0002-3018-4395}}
\email{alberto.hijano@ehu.eus}
\affiliation{Centro de F\'isica de Materiales (CFM-MPC) Centro Mixto CSIC-UPV/EHU, E-20018 Donostia-San Sebasti\'an,  Spain}
\affiliation{Department of Condensed Matter Physics, University of the Basque Country UPV/EHU, 48080 Bilbao, Spain}

\author{Stefan Ili\'{c}}
\email{stefan.ilic@csic.es}
\affiliation{Centro de F\'isica de Materiales (CFM-MPC) Centro Mixto CSIC-UPV/EHU, E-20018 Donostia-San Sebasti\'an,  Spain}

\author{F. Sebasti\'{a}n Bergeret\orcid{0000-0001-6007-4878}}
\email{fs.bergeret@csic.es}
\affiliation{Centro de F\'isica de Materiales (CFM-MPC) Centro Mixto CSIC-UPV/EHU, E-20018 Donostia-San Sebasti\'an,  Spain}
\affiliation{Donostia International Physics Center (DIPC), 20018 Donostia--San Sebasti\'an, Spain}

\begin{abstract}
%We study the anomalous Josephson effect on an Andreev interferometer and show that the anomalous Josephson current can be detected by measuring the oscillations of the dissipative current. For a Josephson junction with spin-filtering barriers we obtain analytic expressions for the Josephson and dissipative currents in different limits and show that their usual and anomalous parts of the current have the same dependence on the orientation of the magnetizations. In a Josephson junction subject to Rashba spin-orbit coupling we demonstrate that the anomalous component of the Josephson and dissipative currents have the same dependence on the spin-orbit coupling strength.\SB{The message has to be more clear. }
Josephson junctions (JJs), where both time-reversal and inversion symmetry are broken,  exhibit a phase shift $\phi_0$ in their current-phase relation. This allows for an anomalous supercurrent to flow in the junction even in the absence of a phase bias between the superconductors.   We show that a finite phase shift also manifests in the so-called Andreev interferometers - a device that consists of a mesoscopic conductor coupled to the $\phi_0$-junction. Due to the proximity effect, the resistance of this conductor is phase-sensitive - it oscillates by varying the phase of the JJ. As a result, the quasiparticle current $I_{\mathrm{qp}}$ flowing through the conductor has an anomalous component, which exists only at finite $\phi_0$. Thus,  the Andreev interferometry could be used to probe the $\phi_0$ effect.
We consider two realizations of the $\phi_0$-junction and calculate $I_{\mathrm{qp}}$ in the interferometer: a superconducting structure with spin-orbit coupling and a system of spin-split superconductors with spin-polarized tunneling barriers.  
\end{abstract}

\maketitle

\section{Introduction} The dc Josephson effect establishes that the current flowing between two superconductors with a phase difference $\phi$, obtained for instance by applying a magnetic flux to the closed circuit, is given as $I_J=I_c\sin{\phi}$. Here $I_c$ is the critical current of the junction. In such junctions the phase-difference of the ground state is $\phi=0$. In a system where (only) time-reversal symmetry is broken, such as  superconductor/ferromagnet/superconductor (S/F/S) structures, it was shown that the current-phase relation can acquire a phase-shift of $\pi$, and therefore such junctions are called $\pi$-junctions ~\cite{Buzdin:1982,Ryazanov:2001,Buzdin:2005,Oboznov:2006}.

In junctions where both time-reversal and inversion symmetries are broken the current-phase relation takes a more general form~\cite{Buzdin:2008}
\begin{equation}\label{I_J}
    I_S=I_c\sin{(\phi+\phi_0)}=I_0^{S}\sin\phi+I_{\mathrm{an}}^{S}\cos\phi\;.
\end{equation}
Such JJs are known as $\phi_0$-junctions by analogy. This effect is referred to as the anomalous Josephson effect (AJE). In general, the current-phase relation of a JJ given by Eq.~\eqref{I_J} can be decomposed into the usual current $I_0^{S}$ and anomalous current $I_{\mathrm{an}}^{S}$. $I_{\mathrm{an}}^{S}$ is non-zero only if the appropriate symmetries are broken, leading to a finite supercurrent even at zero phase difference between the superconductors.

AJE reflects the interplay between spin-dependent fields and superconductivity. This interaction is the basis of several effects and applications that are attracting the interest of a large community, such as topological~\cite{frolov2020topological,sato2017topological,leijnse2012introduction} and unconventional \cite{Bergeret:2005,linder2019odd} superconductivity, superconducting spintronics \cite{Linder:2015}, and novel superconducting electronic elements \cite{ando2020observation}. The most well-known proposals for AJE involve superconducting structures with spin-orbit interaction ~\cite{Reynoso:2008,Buzdin:2008,Zazunov:2009,Brunetti:2013,Yokoyama:2014,Bergeret:2015,Konschelle:2015,Nesterov:2016,Bobkova:2016}, some of which have been successfully tested in experiment \cite{Szombati:2016, Assouline:2019, Strambini:2020,Mayer:2020}. Other theoretical studies have proposed numerous alternative realizations of AJE: in S/F/S junctions with a nonhomogeneous magnetization texture~\cite{Braude:2007,Grein:2009,Margaris:2010,Liu:2010,Kulagina:2014,Moor:2015,Mironov:2015,Silaev:2017}, junctions of unconventional superconductors~\cite{Yip:1995,Sigrist:1998,Kashiwaya:2000,Brydon:2008} and between topologically nontrivial superconducting leads~\cite{Schrade:2017}. An anomalous current-phase relation can also be obtained under non-equilibrium situation in multiterminal structures \cite{dolgirev2018current,Dolgirev:2019current,margineda2021anomalous}.
%Multiple works have shown that it is possible to obtain the AJE in S/F/S structures with a broken magnetization inversion symmetry, such as superconducting structures with spin-orbit interaction~\cite{Reynoso:2008,Buzdin:2008,Zazunov:2009,Brunetti:2013,Yokoyama:2014,Bergeret:2015,Konschelle:2015,Nesterov:2016,Bobkova:2016,Strambini:2020}, conventional S/F/S junctions with a nonhomogeneous magnetization texture~\cite{Braude:2007,Grein:2009,Margaris:2010,Liu:2010,Kulagina:2014,Moor:2015,Mironov:2015,Silaev:2017}, junctions of unconventional superconductors~\cite{Yip:1995,Sigrist:1998,Kashiwaya:2000,Brydon:2008} and between topologically nontrivial superconducting leads~\cite{Schrade:2017}. 
$\phi_0$-junctions could prove to be a key component for quantum electronics, as they  can provide a stable phase bias to quantum circuits, and could therefore be particularly useful in phase-coherent superconducting electronics and spintronics \cite{Linder:2015,Strambini:2020}.

In this work, we consider a $\phi_0$-junction coupled to a mesoscopic conductor, in a device known as Andreev interferometer~\cite{Malshukov:2018,Dolgirev:2019,Volkov:2020,Blasi:2020}. The basic physical idea behind such devices is the following: superconducting correlations are induced in the conductor by the proximity effect, and as a consequence, its resistance becomes sensitive to the phase of the Josephson junction. This means that a simple resistance measurement performed on the conductor can reveal details about the phase dynamics of the adjacent superconducting structure. In the 90's this topic was particularly active, and several types of Andreev interferometers were theoretically proposed~\cite{ZAITSEV:1994,Stoof:1996,Nazarov:1996,Golubov:1997} and experimentally realized~\cite{Petrashov:1993,Petrashov:1995}. Andreev interferometers have been used to study the magnetoresistance oscillations~\cite{Antonov:2001}, electric transport~\cite{Petrashov:2005,Plaszko:2020,Checkley_2011,Volkov:2020,deon2014tuning,galaktionov2012andreev,galaktionov2013current}, and thermopower and thermal transport~\cite{Eom:1998,Parsons:2003,Chandrasekhar_2009,volkov2005long} in S/N structures.

Our  goal  is to establish how the anomalous phase shift $\phi_0$ manifests on the quasi-particle transport through  the Andreev interferometer shown in Fig.~\ref{figure1}(a).    An important advantage of  this geometry is that it allows for a decoupling of the superconducting loop with the $\phi_0$-junction, and the normal wire where the conductance measurement is done, such that the noise associated with the measurement process does not perturb the $\phi_0$ junction.  Our main result is that the  phase-dependent contribution to the dissipative (quasi-particle) current through the vertical arm of the interferometer can be written as:
\begin{equation}
    \delta I_{\mathrm{qp}}(\phi)=I_c^{\mathrm{qp}}\cos(\phi+\phi_0^{\mathrm{qp}})=I^{\mathrm{qp}}_0\cos{\phi}+I^{\mathrm{qp}}_{\mathrm{an}}\sin{\phi}\; .
    \label{eq:qp}
\end{equation}
Therefore, this current also exhibits an anomalous phase shift $\phi_0^{\mathrm{qp}}$. Here $I_0^{\mathrm{qp}}$ is the usual component, which exists in Andreev interferometers with conventional junctions.    $I_{\mathrm{an}}^{\mathrm{qp}}$ is the anomalous component which can only exist in the presence of a $\phi_0$-junction. Note that the phase shift in the Josephson current, $\phi_0$, and in the quasiparticle current, $\phi_0^{\mathrm{qp}}$, are in general not equal, but they have similar magnitude and can be directly related to each other (see the Fig.~\ref{I_qp_final}). Our result 
suggests a way  to experimentally obtain the value of of $\phi_0$ from $\phi_0^{\mathrm{qp}}$ by performing conductance measurements.

We study the two main realizations of $\phi_0$-junctions. Namely, Josephson junctions  with Rashba spin-orbit coupling (SOC) and multilayer ferromagnetic structures [Fig.~\ref{figure1}(b-c)]. In both cases the anomalous phase is related to the existence of a Lifshiftz invariant in the free energy \cite{edelstein_ginzburg_1996,mineev2008nonuniform,bauer_non-centrosymmetric_2012}. In the first example such invariant
%is proportional to the vector product between the magnetic and electric SU(2) fields \cite{Bergeret:2015}. This
stems from an interplay between a Zeeman field and the SOC, whereas in the second example it stems from non-coplanar magnetizations of magnetic layers.
%corresponds to the presence of a Zeeman field perpendicular to the direction of the supercurrent.   In the second example this invariant is determined by the spatial and temporal covariant derivatives of the electric field \cite{baumard2020interplay}, which corresponds to a non-coplanar magnetization of the magnetic layers.} 

%between the noisy measuring loop that include amplifiers, current sources, etc. and the S/N/S structure~\cite{Checkley_2011}. The sensitivity of measurement, the reduced intrinsic noise level of the flux-sensitive loop and the measurement speed makes the Andreev interferometer a promising alternative to superconducting quantum interference devices (SQUIDs), which are the standard tools to perform spectroscopy and study coherent quantum dynamics of circuits.\SB{This last paragraph maybe sounds as copy/paste from old AI papers. We may state here why the AI is important for us. }

%Andreev interferometers have been used to study the magnetoresistance oscillations~\cite{Antonov:2001}, electric transport~\cite{Petrashov:2005,Plaszko:2020,Volkov:2020}, thermopower and thermal transport~\cite{Eom:1998,Parsons:2003,Chandrasekhar_2009} in S/N structures.
\section{The Setup}
\begin{figure}[!t]
    \centering
    \includegraphics[width=0.99\columnwidth]{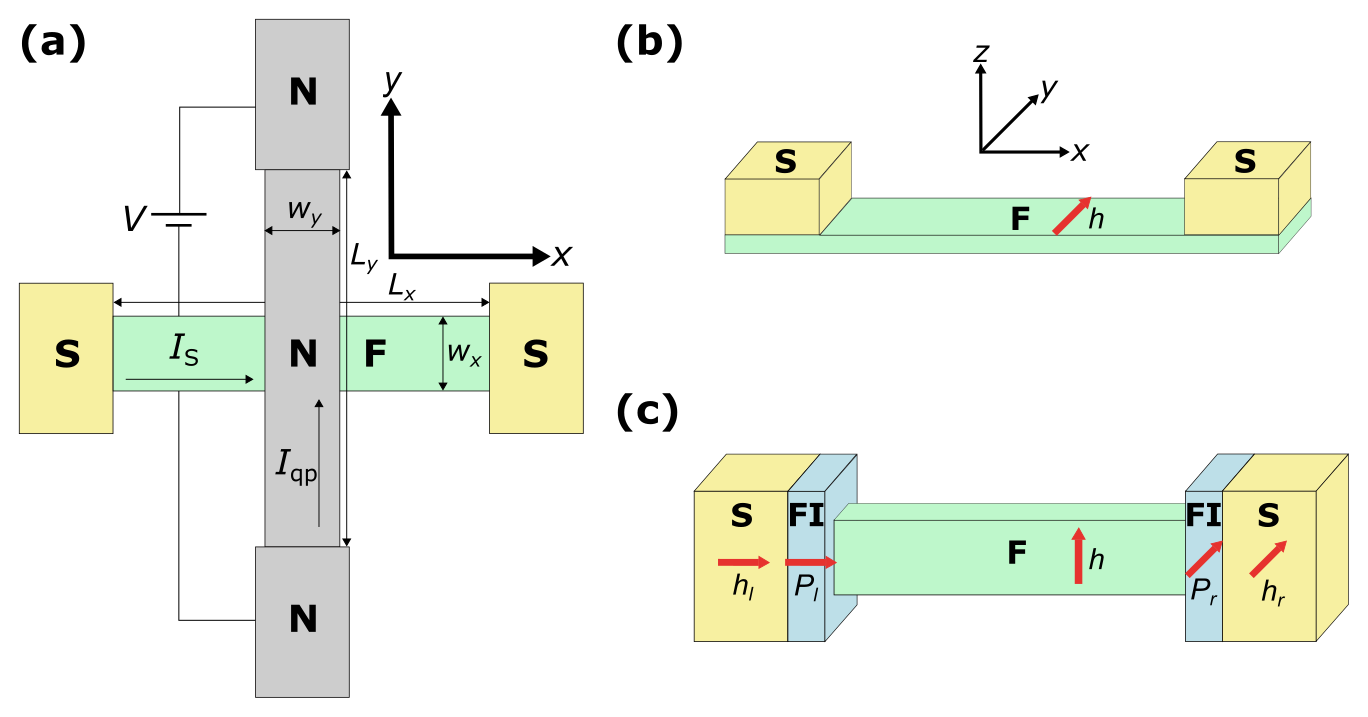}
    \caption{(a) Schematic structure of the Andreev interferometer. (b) S/F/S structure. Here, F is a wire with Rashba spin-orbit coupling, and a spin-splitting field $\boldsymbol{h}$. (c) S/FI/F/FI/S structure. FI layers act as spin-filtering barriers with polarizations $\boldsymbol{P}_{r/l}$, and they induce spin-splitting fields $\boldsymbol{h}_{r/l}$ in adjacent $S$ layers. F is a ferromagnet with an exchange field $\boldsymbol{h}$.}\label{figure1}
\end{figure}

We consider the geometry  shown in Fig.~\ref{figure1}(a). The $\phi_0$ junction lies along the $x$-direction, and consists of a ferromagnetic wire  placed between two superconducting reservoirs. These superconductors are connected in a loop (not shown), so that when a magnetic field is applied through the loop, the resulting flux creates a phase difference between them and leads to a Josephson current flowing along the $x$ wire. 
An additional  normal wire (N) is placed perpendicularly to the F wire  (on the $y$ direction). N is connected to  two normal reservoirs. We assume that the F and N wires  intersect at their midpoints.  
A voltage difference between the normal electrodes leads to a quasiparticle current $I_{\mathrm{qp}}$ in the $y$ wire which can be decomposed in two contributions: {$I_{\mathrm{qp}}=I_\Omega+\delta I_{\mathrm{qp}}(\phi)$, where $I_\Omega$ is the usual Ohmic contribution, whereas $\delta I_{\mathrm{qp}}(\phi)$ is the phase-dependent part given in Eq. (\ref{eq:qp}}). %\textcolor{blue}{[here we renormalize $I_\Omega$ with the constant component of $\delta I_{\mathrm{qp}}(\phi)$ and only consider the components which depend explicitly on the phase $\phi$]}.
%\textcolor{green}{[SI: The sentence in the brackets is confusing. Is it necessary? We can maybe simply say that $I_\Omega$ is the phase-indepent part. ]}
The latter  is affected by the proximity effect with the $x$ wire. %Namely it depends on the phase difference between the S electrodes:

In the  rest of this work we determine the usual and anomalous components of $\delta I_{\mathrm{qp}}(\phi)$  for two different realizations of a $\phi_0$-junction: a S/F/S junction with Rashba SOC [Fig.~\ref{figure1}(b), Sec.~\ref{Sec3}] and a S/FI/F/FI/S junction, where FI stands for a ferromagnetic insulator~[Fig.~\ref{figure1}(c), Sec.~\ref{Sec4}]. In the first example, a magnetic field and a spin-orbit coupling provide time-reversal and inversion symmetry breaking, respectively, which leads to the anomalous phase shift \cite{Buzdin:2008, Bergeret:2015}. Here the anomalous Josephson current is $I_{\mathrm{an}}^S \propto h \kappa_\alpha$, with $\kappa_\alpha$ being a parameter associated with singlet-triplet conversion due to the SOC, and $h$ is a weak exchange or Zeeman  field [see Eq.~\eqref{Usadel Rashba}]. In the second example, the $\phi_0$ effect occurs if the FI tunneling barriers are spin-polarized, so that the barrier polarizations $\boldsymbol{P}_{r/l}$ and the magnetization direction in the F layer $\boldsymbol{h}$ are non-coplanar, and therefore the magnetization inversion symmetry is broken \cite{Braude:2007,Grein:2009,Margaris:2010,Liu:2010,Kulagina:2014,Moor:2015,Mironov:2015,Silaev:2017}. The anomalous Josephson current is $I_{\mathrm{an}}^S\propto \chi$, where $\chi=\boldsymbol{n}_h\cdot(\boldsymbol{n}_l \times \boldsymbol{n}_r)$ is the so-called chirality of the junction (see Fig.~\ref{figure1}). Here, $\boldsymbol{n}_{h}$ and $\boldsymbol{n}_{r/l}$ are the unit vectors along the exchange field and polarization directions. We will show that in both examples, the interferometer quasiparticle anomalous current $I_{\mathrm{an}}^{\mathrm{qp}}$ has the same  dependence on the spin-dependent fields as 
the anomalous Josephson current $I_{\mathrm{an}}^{S}$. %(similar holds for the usual currents $I_0^S$ and $I_0^{\mathrm{qp}}$). 
Namely $I_{\mathrm{an}}^{\mathrm{qp}}\propto h \kappa_\alpha$ in the first example, and $I_{\mathrm{an}}^{\mathrm{qp}} \propto \chi$ in the second example.
%\textcolor{blue}{[MORE REFERENCES to the latter example.  AND EXPLAIN THE MAIN RESULT: i.e. how the ANOMALOUS CURRENT DEPENDS ON THE CHIRALITY PARAMETER AND THE SOC*By component.]}

\section{Josephson junction with Rashba SOC \label{Sec3}}
We first study a S/F/S structure, shown in Fig.~\ref{figure1}(b). Here, F is a wire with Rashba spin-orbit coupling (SOC), and an exchange field $\boldsymbol{h}$ which comes either from intrinsic magnetization (F is a ferromagnet) or from an externally applied magnetic field. For this configuration, the anomalous Josephson current is only affected by the component of the exchange field perpendicular to the current direction $x$, so we consider a field oriented along the $y$ direction $\boldsymbol{h}=h \boldsymbol{y}$ in order to maximize the $\phi_0$ effect \cite{Bergeret:2015}.

%We analyze the system using the Green's function (GF) technique. The GF is obtained from the quasiclassical equations, which in the dirty limit reduce to a diffusive-like equation, known as the Usadel equation~\cite{Usadel}. To study a system with spin-orbit coupling, one should use the covariant Usadel equation~\cite{Tokatly:2017}. We assume that the proximity effect on the wires due to the electrodes is weak due to a small S/F transmission coefficient. In this case, the linearized Usadel equation takes the form

We describe the system using the quasiclassical Green's function (GF) formalism \cite{belzig1999quasiclassical}. In the diffusive limit, GFs are obtained as a solution of the Usadel equation \cite{Usadel}.  SOC can be  included as a background SU(2) field \cite{bergeret2013singlet,bergeret2014spin,Tokatly:2017}.
Superconducting correlations are described by the condensate GF, $\hat f$, which is a $2 \times 2$ matrix in spin space that consists of a singlet component, $f_0$ and, in general, three triplet components, $f_j\sigma_j$, where $j=1,2,3$ and $\sigma_j$ are the three Pauli matrices.
We assume that the proximity effect in the F wire is weak  due, for example,  to low  S/F transmission coefficient. In this case, the  Usadel equation can be linearized \cite{Bergeret:2015}. For the situation under consideration, transport in $x$-direction and $h$-field in $y$-direction, only the condensate components $f_0$ and $f_2$ are finite and  satisfy:  
\begin{subequations}\label{Usadel Rashba}
\begin{align}
	\pm\partial^2_{xx} f^{R/A}_0+i\kappa_\epsilon^2 f^{R/A}_0 -i\kappa^2_F f^{R/A}_2 -\kappa_\alpha\partial_x f^{R/A}_2 =& 0\\
	\pm\partial^2_{xx} f^{R/A}_2+i\kappa_\epsilon^2 f^{R/A}_2 -i\kappa^2_F f^{R/A}_0 
	-\kappa_\alpha\partial_x f^{R/A}_0 =& 0
\end{align}
\end{subequations}
where $\kappa_\epsilon^2=2\epsilon/D$, $\kappa_F^2=2h/D$ and $\kappa_\alpha=4\alpha^3\tau/m$. Here, $\epsilon$ is the energy, $D$ is the diffusion constant, $h$ is the exchange field and $\alpha$ is the Rashba coupling constant. The upper and lower sign correspond to the retarded and advanced condensate GFs $\hat{f}^{R/A}$ respectively. In the following we omit the superscript to simplify the notation. Moreover,  to simplify the calculation, in Eq.~\eqref{Usadel Rashba} we have neglected the renormalization of the exchange field by the SOC, and the  relaxation of the triplet component due to SOC~\cite{Strambini:2020}. 

%\SB{REMARK: it seems that we have in the manuscript  two different notations for the same object: $f_y=f_2$? }

The Usadel equation~\eqref{Usadel Rashba} is supplemented by boundary conditions describing the interfaces between different materials. The S/F junctions are described by the generalized  Kuprianov-Lukichev conditions~\cite{Kuprianov-Lukichev}
\begin{subequations}\label{BC_Rashba_components}
\begin{align}
	\pm\partial_n f_{0,r/l}+\eta_{r/l}\kappa_\alpha f_{2,r/l} =& \mp\frac{1}{\gamma} \mathcal{F}_0 e^{i \eta_{r/l}\phi/2}\\
	\partial_n f_{2,r/l} =& 0\; .
\end{align}
\end{subequations}
Here, $\mathcal{F}_0=\Delta/\sqrt{\Delta^2-\epsilon^2}$ is the anomalous GF of the superconducting electrode, $\partial_n$ is the normal derivative at the surface and $\gamma=\sigma_F R_b$ is the parameter describing the barrier strength, where $R_b$ is the normal-state tunneling resistance per unit area and $\sigma_F$ is the conductivity of the ferromagnet. $\eta_{r/l}=\pm 1$ for the right ($x=L_x/2$) and left boundaries ($x=-L_x/2$).

%The S electrodes do not show spin-splitting, so $F_S$ is given by the BCS GF.

%The $y$ wire is a normal wire, so the exchange field is equal to $\boldsymbol{h}=0$.

%so the current along the $y$ wire carries information from the $x$ wire

The condensate function in the $y$ wire, $\hat{f}_{y}$, is induced by the proximity effect with the $x$ wire. To find $\hat{f}_y$, we start from the Kuprianov-Lukichev condition describing the interface between the two wires, and the Usadel equation in the $y$-wire. Provided that the widths of the wires $w_{x,y}$ are much smaller than the superconducting coherence length, we can integrate the Usadel equation over the cross-section of the wire.  If the interface resistance is much larger than the resistance of the wires, $R_B \gg L_{x,y}/\sigma_{F,N}$, we find the equation determining $\hat{f}_y$: 
\begin{equation}\label{Usadely}
	\pm\partial_{yy}^2\hat{f}_{y}+i\left.\kappa'_\epsilon\right.^2\hat{f}_{y}=-\frac{w_x}{\gamma_0^2}\hat{f}(0)\delta(y)\; .
\end{equation}
%\SB{here i have a problem with the notation. I think we have not defined $f_x$. Such that we can write $\hat{f}(0)$ right?}
Here, $\gamma_0^2=R_B \sigma_N w_y$ and $\left.\kappa'_\epsilon\right.^2=2\epsilon/D_y$, with $R_B$ being the resistance per unit area of the interface of the $x$ and $y$ wires and $\sigma_N$ is the normal-state conductance of the $y$ wire. The Dirac delta term describes the proximity effect, and is a source term. The contact of the $y$ wire with the  normal reservoirs is assumed to be ideal so that the condensate functions vanish  at the ends of the wire is $\hat{f}_{y}(\pm L_y/2)=0$.

%No voltage is applied on the S/F/S wire, so the current is given by the condensate current. 
A voltage bias $V$ is applied between the normal electrodes.  Due to  our assumption of large $R_B$  we can neglect the  inverse proximity effect.
%\SB{What is the small  dimensionless parameter that allows such an assumption?} \textcolor{blue}{[SI:Done]}
Thus, in leading order the phase-dependent correction to the quasi-particle current is  given by \cite{Volkov1996,Volkov1998methods,volkov1998phase}
\begin{equation}\label{quasiparticle_current}
	\delta I_{\mathrm{qp}}=\frac{-\sigma_N}{16 e L_y}\int d\epsilon F_T(\epsilon,V/2)\left\langle\mathrm{Tr}(\check{f}^R_{y}-\check{f}^A_{y})^2\right\rangle\; .
\end{equation}
Here $\langle...\rangle=1/L_y \int_{-L_y/2}^{L_y/2}dy\, (...)$ denotes average over the length, $\check{f}^{R/A}$ is the $4 \times 4$ matrix GF in Nambu-spin space [see Eq.~\eqref{f Nambu-spin} in the Appendix] and $F_T$ is  defined as $F_T(\epsilon,V)=\frac{1}{2}[\tanh \frac{\epsilon+eV}{2T}-\tanh \frac{\epsilon-eV}{2T}]$.
Solving the boundary value problem, Eqs.~\eqref{Usadel Rashba} and~\eqref{BC_Rashba_components}, we first calculate the $\hat{f}$ for the $x$-wire, and then $\hat{f}_y$ for the $y$-wire from Eq.~\eqref{Usadely}. Using Eq.~\eqref{quasiparticle_current} we then obtain the usual and anomalous quasiparticle currents entering Eq.~\eqref{eq:qp}.

\begin{figure*}[!t]
    \centering
    \includegraphics[width=0.99\textwidth]{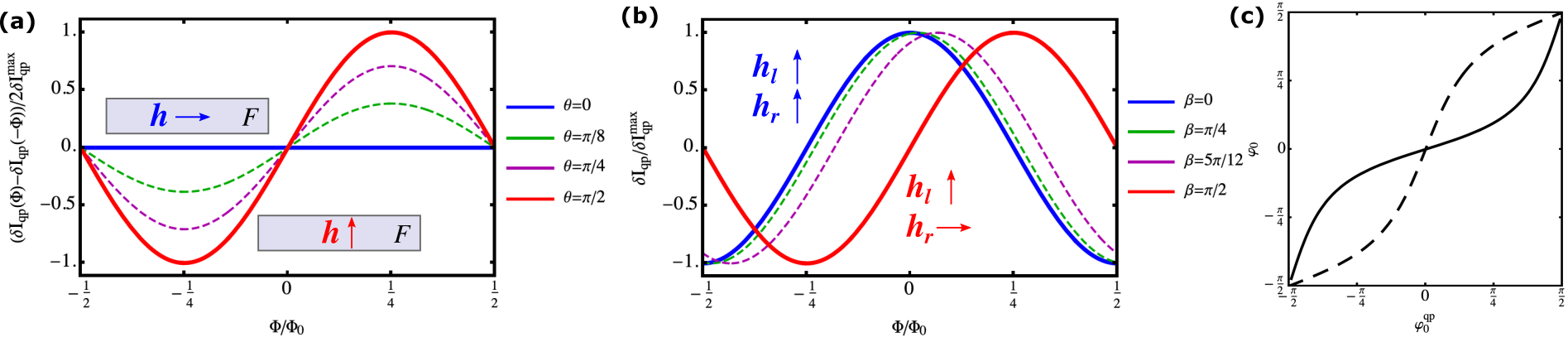}
    \caption{(a) Component of the quasiparticle current odd in flux, which corresponds to $I^{\mathrm{qp}}_{\mathrm{an}}$, for the configuration shown in Fig~\ref{figure1}(b). $\theta$ is an angle between the in-plane exchange field in the F layer and the $x$-axis. The Josephson phase is given by $\phi=2\pi\Phi/\Phi_0$, where $\Phi$ is the applied flux and $\Phi_0$ is the flux quantum. (b) Quasiparticle current along the $y$ wire for different magnetization directions for the configuration shown in Fig~\ref{figure1}(c). The exchange fields on the S electrodes $\boldsymbol{h}_{r/l}$ are taken to be perpendicular to the exchange field on the F wire in order to maximize the current, while they form an angle $\beta$. (c) Relation between the anomalous phase-shifts in the Josephon current, $\phi_0$, and in the quasiparticle current, $\phi_0^{\mathrm{qp}}$, for the junction with Rashba SOC (solid) and S/FI/F/FI/S junction (dashed), calculated from the expressions provided in Appendices~\ref{sec1} and~\ref{sec2}, respectively. In the first case, the value of the exchange field is $h=0.1\Delta$ and the Rashba coupling constant ranges from $\kappa_\alpha \xi_0 \in [-0.5,0.5]$. In the second case, the exchange field is $h=0.1\Delta$ and the splitting of the superconducting electrodes are $h_{r/l}=0.1\Delta$, where they form an angle $\beta \in [-\pi/2,\pi/2]$. The temperature is $T=0.01 T_{c0}$ in both cases.}\label{I_qp_final}
\end{figure*}

Up to the leading order terms in exchange field and Rashba SOC, the quasiparticle current takes the following form:
\begin{align}
    \label{I^qp_0_2}
	I_0^{\mathrm{qp}} &= c_1,\\
	\label{I^qp_an_2}
	I_{\mathrm{an}}^{\mathrm{qp}} &= c_2 h \kappa_\alpha\; .
\end{align}
The factors $c_{1}$, and $c_2$ depend on  $T$ and $L_{x,y}$, and their exact form is given by Eqs.~\eqref{Iy_0_Rashba} and~\eqref{Iy_an_Rashba} in the Appendix.
%\SB{$c_1'$ interference term[CITATIONS]}
%Studying the amplitude of the anomalous Josephson current [see Supplementary Information], we see that we need both a finite exchange field and Rashba SOC strength to obtain an anomalous quasiparticle current.
Both components of the quasiparticle current depend on the spin-dependent fields in the same way as the components of Josephson current \cite{Bergeret:2015}: $I_0^S$ and $I_0^{\mathrm{qp}}$ are independent of these fields, whereas $I_{\mathrm{an}}^S,I_{\mathrm{an}}^{\mathrm{qp}}\sim h \kappa_\alpha$.
Note that the result for the anomalous current, Eq.~\eqref{I^qp_an_2}, also holds in the case when exchange field in the F layer is not fully aligned with the $y$-direction, by taking $h=h_{tot} \sin \theta$. Here $h_{tot}$ is an arbitrarily oriented in-plane field, and $\theta$ is an angle between the field and the $x$-direction. To illustrate this, in Fig.~\ref{I_qp_final}(a) we plot the odd component of the anomalous quasiparticle current, $I_{\mathrm{an}}^{\mathrm{qp}}\sin \phi=\frac{1}{2}[I_{\mathrm{qp}}(\phi)-I_{\mathrm{qp}}(-\phi)]$, for different values of $\theta$. The current is normalized with respect to its maximum value for clarity. For $\theta=0$, the exchange field is parallel to the wire, so there is no anomalous phase shift and $I^{\mathrm{qp}}_{\mathrm{an}}$ vanishes. The $\phi_0$ effect is maximized for $\theta=\pi/2$, where the exchange field is perpendicular to the wire.
In Fig.~\ref{I_qp_final}(c) we  plot the relation between the $\phi_0$ phase-shift in the Josephson current, and the phase-shift measured in the quasiparticle current $\phi_0^{\mathrm{qp}}$. All  expressions presented in this work are valid for arbitrary temperature. For the numerical computations however, we only  focus on low temperatures, $T=0.01 T_{c0}$,  where  the magnitude of the quasiparticle current is maximized. %\textcolor{green}{[SI: The specific temperature can be written in the caption of the figure, no need in the main text.]}

\section{S/FI/F/FI/S junction \label{Sec4}}
Another configuration to obtain a $\phi_0$-junction is a S/FI/F/FI/S junction with non-coplanar magnetizations [Fig.~\ref{figure1}(c)]. This configuration has not yet been realized in experiment, but has been theoretically predicted to show AJE~\cite{Braude:2007,Grein:2009,Margaris:2010,Liu:2010,Kulagina:2014,Moor:2015,Mironov:2015,Silaev:2017}. In these structures, the role of the FI layers is two-fold: firstly, they induce an exchange field $\boldsymbol{h}_{r/l}$ in the adjacent S layer, and secondly, they act as spin-polarized tunneling barriers with a polarization  $\boldsymbol{P}_{r/l}$.
The linearized Usadel equation in the F layer reads
\begin{equation}\label{Usadel linearized}
	\pm\partial_{xx}^2\hat{f}+i\kappa_\epsilon^2\hat{f}-i\frac{\kappa_F^2}{2}\{\sigma_3,\hat{f}\}=0\; ,
\end{equation}
where $\{.,.\}$ is an anticommutator. We have assumed, without loss of generality, that the exchange field in the F-wire points on the $\boldsymbol{z}$ direction.

The S/F junctions with spin-filtering barriers are described by the generalized Kuprianov-Lukichev boundary condition~\cite{Bergeret:2012,Eschrig:2015}. The exchange fields $\boldsymbol{h}_{r/l}$ induced via the magnetic proximity effect in the S-electrodes point in the same direction as  the polarization vectors $\boldsymbol{P}_{r/l}$.   The linearized boundary conditions read
\begin{multline}\label{boundary_condition_linearized}
	\pm \gamma \partial_n \hat{f}_{r/l}=\frac{1}{2}[\hat{\mathcal{G}}_{r/l} \boldsymbol{P}_{r/l}\cdot\boldsymbol{\sigma},\hat{f}_{r/l}] \\ + \frac{1}{2}\{\hat{\mathcal{G}}_{r/l},\hat{f}_{r/l}\} \mp \sqrt{1-P_{r/l}^2}\hat{\mathcal{F}}_{r/l} e^{i \eta_{r/l}\phi/2} \; .
\end{multline}
Here, $\hat{\mathcal{G}}_{r/l}$ and $\hat{\mathcal{F}}_{r/l}$ are the normal and anomalous GFs of the spin-split superconducting electrode, respectively. In the weak exchange field limit, they are given by 
\begin{subequations}\label{GF S electrodes}
\begin{align}
    \hat{\mathcal{G}}_{r/l} &= \mathcal{G}_0 -\boldsymbol{h}_{r/l}\cdot\boldsymbol{\sigma}\frac{d\mathcal{G}_0}{d\epsilon}\\
    \hat{\mathcal{F}}_{r/l} &= \mathcal{F}_0 -\boldsymbol{h}_{r/l}\cdot\boldsymbol{\sigma}\frac{d\mathcal{F}_0}{d\epsilon}
\end{align}
\end{subequations}
with $\mathcal{G}_0=-i\epsilon/\sqrt{\Delta^2-\epsilon^2}$.
%\textcolor{blue}{[Comment: Write explicity $\hat{\mathcal{G}}_S$ and $\hat{\mathcal{F}}_S$ for weak $h_{r/l}$. Also, in the boundary condition \eqref{boundary_condition_linearized}, write indices $l$ and $r$ to distinguish different interfaces.]}

Generally, the coherence length in the ferromagnetic layer is much shorter than in a normal metal $\kappa_F^{-1} \ll \kappa_N^{-1}$, where $\kappa_N=\sqrt{2T/D}$. We assume the long-junction regime, so that $\kappa_F^{-1} \ll L_x \ll \kappa_N^{-1}$. In this regime the length of the $x$ wire $L_x$ is much longer than the penetration length of the Cooper pairs $\kappa_F^{-1}$ in the F layer, so that the condensate functions $\hat{f}$ and $\hat{f}_y$ are  mediated primarily by the long-range triplet superconducting correlations~\cite{Bergeret:2005,Buzdin:2011,Samokhvalov:2019}, whereas the singlet and short-range triplet correlations decay over the length $\kappa_F^{-1}$.
%\textcolor{blue}{[add citations]}

To calculate the interferometer current, we proceed similarly as in the previous example. First, from Eqs.~\eqref{Usadel linearized} and \eqref{boundary_condition_linearized}, we find $\hat{f}$ in the $x$-wire, and then calculate the current in the $y$-wire from Eqs.~\eqref{Usadely} and \eqref{quasiparticle_current}. The usual and anomalous quasiparticle currents are given by~\cite{Silaev:2017}
\begin{align}
    \label{I^qp_0_1}
	I^{\mathrm{qp}}_0 &= c_3 \sqrt{1-P_r^2}\sqrt{1-P_l^2}\gamma^{-2}\boldsymbol{h}_{l\perp}\cdot\boldsymbol{h}_{r\perp}\\
	\label{I^qp_an_1}
	I^{\mathrm{qp}}_\mathrm{an} &= c_4 \sqrt{1-P_r^2}\sqrt{1-P_l^2}\gamma^{-3}(P_l h_r +P_r h_l)\boldsymbol{z}\cdot(\boldsymbol{n}_l \times \boldsymbol{n}_r)\; ,
\end{align}
where $\boldsymbol{h}_{r/l\perp}=\boldsymbol{h}_{r/l}-(\boldsymbol{h}_{r/l}\cdot\boldsymbol{h})\boldsymbol{h}/h^2$ are the components of $\boldsymbol{h}_{r/l}$ perpendicular to $\boldsymbol{h}$ [see Eqs.~\eqref{c_3} and~\eqref{c_4} in the Appendix for the exact form of the coefficients $c_3$ and $c_4$]. 

From Eq.~\eqref{I^qp_an_1}, we see that the anomalous quasiparticle current is proportional to the scalar triple product of the magnetizations $I^{\mathrm{qp}}_{\mathrm{an}} \propto \chi=\boldsymbol{z}\cdot(\boldsymbol{n}_l \times \boldsymbol{n}_r)$ \cite{Silaev:2017}. Here, $\chi$ is the junction chirality, and it is non-zero only if the barrier polarizations and the magnetization direction are non-coplanar. 
As in the previous example, the quasiparticle current and the Josephson current have the same dependence   on the spin-dependent fields:  for the usual components  $I_0^S, I_0^{\mathrm{qp}}\propto \boldsymbol{h}_{l\perp}\cdot\boldsymbol{h}_{r\perp}$, and for the anomalous components $I_{\mathrm{an}}^S, I_{\mathrm{an}}^{\mathrm{qp}}\propto \chi$.

Usually, $I_0^{\mathrm{qp}}$ is the dominant contribution to the interferomenter current, as it is of the lower order in the small barrier parameter $\gamma^{-1}$, namely $I_{\mathrm{an}}^{\mathrm{qp}}/I_{0}^{\mathrm{qp}}$,$I_{\mathrm{an}}^{S}/I_{0}^{S} \sim \gamma^{-1} \ll 1$.
However,  if  $\boldsymbol{h}_{l\perp}\cdot\boldsymbol{h}_{r\perp}=0$ the usual component vanishes and only the  anomalous current  contributes. In other words,  the measured quasiparticle current is directly linked to the $\phi_0$ effect. To illustrate this point further, in Fig.~\ref{I_qp_final}(b) we use the analytical formulas (\ref{c_3}-\ref{c_4}) to plot the normalized quasiparticle current for different magnetic configurations. Here, $\beta$ is the angle formed by the exchange fields on the S electrodes $\boldsymbol{h}_{r/l}$, which are taken to be perpendicular to $\boldsymbol{h}$. Unlike in panel (a), here we plot the total anomalous quasiparticle current $I_{\mathrm{an}}^{\mathrm{qp}}$ to stress the shift from an even-in-phase to odd-in-phase behavior when the angle between the electrode magnetizations increases. For $\beta=0$, there is no $\phi_0$ effect, so the current is even in the phase. $I_0^{\mathrm{qp}}$ decreases with increasing $\beta$ and vanishes for $\beta=\pi/2$. In this case, the oscillation of $\delta I_{\mathrm{qp}}$ are given by the anomalous quasiparticle current, so that $\delta I_{\mathrm{qp}}$ becomes odd in $\phi$.
In Fig.~\ref{I_qp_final}(c) we  plot the relation between the $\phi_0$ and $\phi_0^{\mathrm{qp}}.$

It is worth mentioning, that to simplify equations and  obtain analytical solutions,  we have assumed a low barrier transmission at the S/F interfaces and between the $x$ and $y$ wires. Consequently, the obtained quasiparticle current, being proportional to powers of a small interface parameter,  is  also small. However, the  findings of our work should  still hold qualitatively in  setups with smaller interface resistances, where the quasiparticle currents should be significantly larger.
In other words,  our results give a lower bound of the current amplitude. Moreover, the expressions for the quasiparticle current, Eqs. (\ref{I^qp_0_2}-\ref{I^qp_an_2}, \ref{I^qp_0_1}-\ref{I^qp_an_1})  are valid for all  temperatures. Indeed, the temperature dependence enters 
the coefficients $c_i$, ($i=1,2,3,4$),  given in  Eqs.~(\ref{Iy_0_Rashba}-\ref{Iy_an_Rashba},\ref{c_3}-\ref{c_4}).  From these expressions one can show  that the quasiparticle current amplitudes are maximized at low temperatures, and they decrease monotonically towards zero at the superconducting critical temperature. This behaviour of $\delta I^{\mathrm{qp}}$ is in contrast with the temperature behaviour of the critical Josephson current, whose sign may be reversed by changing the temperature \cite{Ryazanov:2001} ({\it i.e.} when a $0$-$\pi$ transition occurs).

\section{Conclusion}
In summary, we have studied the current-phase relation of an Andreev interferometer with an anomalous Josephson junction. We have shown how the quasiparticle current through the normal arm of the interferometer is affected by the appearance of an anomalous phase $\phi_0$. Specifically, we have studied the AJE in S/F/S structures with spin-splitting field and Rashba SOC or spin-filtering barriers. Our results show that there is also an anomalous contribution to the phase-dependent part of the quasiparticle current proportional to $\sin \phi$ when $\phi_0$ is different from $0$ and $\pi$. Moreover, the usual and anomalous quasiparticle currents have the same dependence on the spin-dependent fields as the anomalous Josephson current. Suitable  materials for the realization of the anomalous Andreev interferometer are InSb~\cite{Szombati:2016}, Bi wires \cite{murani2017ballistic}, Bi$_2$Se$_3$~\cite{Assouline:2019} and InAs~\cite{Mayer:2020,baumgartner2021josephson} due to the large spin-orbit coupling, in combination with conventional superconductors and  normal metals. In particular, in systems with InAs a large phase shift $\phi_0\approx \pi/2$ was experimentally observed \cite{Strambini:2020} in the Josephson current, and based on our findings, we expect equally strong effect in the Andreev interferometer geometry. For the ferromagnetic interferometers we propose EuS/Al structures to engineer a $\phi_0$-junction in a S/FI/F/FI/S junction due to the well-defined splitting and strong barrier polarization~\cite{Rouco:2019}, while the F layer can consist of a Co wire~\cite{wang2010interplay,Khaire:2010}.

\section*{Acknowledgements}  This work was partially  funded by the Spanish Ministerio de Ciencia, Innovación y Universidades (MICINN) through Projects FIS2017-82804-P   and 
PID2020-114252GB-I00 (SPIRIT), and EU’s Horizon 2020 research and innovation program under Grant Agreement No. 800923 (SUPERTED). A.H. acknowledges funding by the University of the Basque Country (Project PIF20/05).

%In summary, we have studied the current-phase relation on an Andreev interferometer and we have shown that the current along the wire perpendicular to the Josephson junction is affected by the appearance of an anomalous phase $\phi_0$. Specifically, we have studied the AJE in S/F/S structures with spin-splitting field and Rashba SOC or spin-filtering barriers. Analytical results in various limits show that the quasiparticle current on the perpendicular wire acquires an anomalous quasiparticle current which is proportional to $\sin{\phi}$ when $\phi_0$ is non-trivial. Moreover, the usual and anomalous quasiparticle currents have the same SOC strength/field dependence as the Josephson current. Some suitable materials for the experimental measurement of the anomalous Josephson current in an Andreev interferometer are InSb~\cite{Szombati:2016}, Bi$_2$Se$_3$~\cite{Assouline:2019} and InAs~\cite{Mayer:2020} due to the large spin-orbit coupling and g-factors. EuS/Al is an ideal material combination to engineer a $\phi_0$-junction in a S/FI/F/FI/S structure due to the well-defined splitting and strong barrier polarization~\cite{Rouco:2019}, while the F layer can be consist of Co~\cite{Khaire:2010}.

\onecolumngrid

%\begin{appendices}
\appendix
\numberwithin{equation}{section}
\renewcommand{\thesubsection}{\arabic{subsection}}
\section{Josephson junction with Rashba SOC \label{sec1}}

In this Appendix we present a detailed derivation of the expressions used in the main text for the currents in the presence  of Rashba SOC. 
In Sec.~\ref{subsec1A} we first present the derivation of the Josephson current in the $x$-wire, followed by the derivation of the quasiparticle current in the $y$-wire in Sec.~\ref{subsec1b}.

%In this section we derive the equations and the results presented in the ``Josephson junction with Rashba SOC'' section of the main text. We first describe the system using the quasiclassical Green's function (GF) formalism and solve the Usadel equation to obtain the supercurrent in the $x$ wire and the quasiparticle current in the $y$ wire (see Fig. 1a in the main text).

\subsection{Current along the \texorpdfstring{$x$}{} wire \label{subsec1A}}
%A possible configuration to obtain the $\phi_0$ effect is the presence of a spin-splitting field and a spin-orbit coupling (SOC) on a S/F/S structure (see Fig. 1(b) in the main text). We assume a SO coupling of the Rashba type; for this configuration, the anomalous Josephson current is only affected by the component of the exchange field perpendicular to the current direction $x$, so we consider a field oriented along the $y$ direction $\boldsymbol{h}=h \boldsymbol{y}$ in order to maximize the $\phi_0$ effect~\cite{Bergeret:2015}.

We start by solving the linearized Usadel equation, Eq.~\eqref{Usadel Rashba}, with the appropriate boundary conditions - Eq.~\eqref{BC_Rashba_components}. First, let us note that the retarded and advanced anomalous Green's functions (GFs) in Nambu-spin space have the following structure
\begin{equation}\label{f Nambu-spin}
	\check{f}=
	\begin{pmatrix}
		0 & \hat{f}(\epsilon)\\
		\overline{\hat{f}(-\epsilon)} & 0
	\end{pmatrix}\; ,
\end{equation}
where $\hat{\bar{X}}=\mathcal{T}\hat{X}\mathcal{T}^{-1}$, and $\mathcal{T}=i \sigma_2 \mathcal{K}$ is the time-reversal transformation, with $\mathcal{K}$ being the complex conjugate operation. Moreover, we can relate $\check{f}^A$ to $\check{f}^R$ as
\begin{equation}
	\check{f}^A(\epsilon,\boldsymbol{h},\alpha)=\check{f}^R(-\epsilon,-\boldsymbol{h},-\alpha)\; .
\end{equation}
In the following we only write the retarded GF and omit the superscript to simplify the notation.
%The linearized Usadel equation and boundary conditions for $\hat{f}$ are given by
%Main text cross-reference
%Eqs.~(3) and (4) in the main text.
%In order to simplify the calculation, we neglect the renormalization of the exchange field by the SOC, and the relaxation of the triplet component due to SOC~\cite{Strambini:2020}. In addition, we assume that the SOC strength is weak ($\kappa_\alpha L \ll 1$ and $\kappa_\alpha/\kappa_\epsilon \ll 1$) and treat the Rashba term perturbatively. The triplet components $f_1$ and $f_3$ are equal to zero since the exchange field and the effective magnetic field due to the Rashba interaction lie on the $y$ direction.

We find the condensate function $\hat{f}$ perturbatively in $\kappa_\alpha$, keeping terms up to the first order in this parameter. The solution is
\begin{align}
    \begin{split}
	f_0 &=\left[\left(\left(1+\frac{\kappa_\alpha x}{2}\right)A_{1,+}+B_{1,+}\right)e^{i\phi/2}+\left(\left(1+\frac{\kappa_\alpha x}{2}\right)A_{2,+}+B_{2,+}\right)e^{-i\phi/2}\right]e^{\kappa_+ x}\\
	&+\left[\left(\left(1+\frac{\kappa_\alpha x}{2}\right)A_{2,+}-B_{2,+}\right)e^{i\phi/2}+\left(\left(1+\frac{\kappa_\alpha x}{2}\right)A_{1,+}-B_{1,+}\right)e^{-i\phi/2}\right] e^{-\kappa_+ x}\\
	&+\left[\left(\left(1-\frac{\kappa_\alpha x}{2}\right)A_{1,-}-B_{1,-}\right)e^{i\phi/2}+\left(\left(1-\frac{\kappa_\alpha x}{2}\right)A_{2,-}-B_{2,-}\right)e^{-i\phi/2}\right] e^{\kappa_- x}\\
	&+\left[\left(\left(1-\frac{\kappa_\alpha x}{2}\right)A_{2,-}+B_{2,-}\right)e^{i\phi/2}+\left(\left(1-\frac{\kappa_\alpha x}{2}\right)A_{1,-}+B_{1,-}\right)e^{-i\phi/2}\right] e^{-\kappa_- x},
    \end{split}
    \label{eqGF0}
   \end{align}
 \begin{align}
 	\begin{split}
	f_2 &=\left[\left(\left(1+\frac{\kappa_\alpha x}{2}\right)A_{1,+}+B_{1,+}\right)e^{i\phi/2}+\left(\left(1+\frac{\kappa_\alpha x}{2}\right)A_{2,+}+B_{2,+}\right)e^{-i\phi/2}\right]e^{\kappa_+ x}\\
	&+\left[\left(\left(1+\frac{\kappa_\alpha x}{2}\right)A_{2,+}-B_{2,+}\right)e^{i\phi/2}+\left(\left(1+\frac{\kappa_\alpha x}{2}\right)A_{1,+}-B_{1,+}\right)e^{-i\phi/2}\right] e^{-\kappa_+ x}\\
	&-\left[\left(\left(1-\frac{\kappa_\alpha x}{2}\right)A_{1,-}-B_{1,-}\right)e^{i\phi/2}+\left(\left(1-\frac{\kappa_\alpha x}{2}\right)A_{2,-}-B_{2,-}\right)e^{-i\phi/2}\right] e^{\kappa_- x}\\
	&-\left[\left(\left(1-\frac{\kappa_\alpha x}{2}\right)A_{2,-}+B_{2,-}\right)e^{i\phi/2}+\left(\left(1-\frac{\kappa_\alpha x}{2}\right)A_{1,-}+B_{1,-}\right)e^{-i\phi/2}\right] e^{-\kappa_- x}\; ,
    \end{split}
    \label{eqGF2}
\end{align}

where $\kappa_\pm=\sqrt{-i\kappa_\epsilon^2 \pm i \kappa_F^2}$, and the coefficients are given by
\begin{subequations}
\begin{align}
	A_{1,\pm} &=\frac{\mathcal{F}_0}{4\gamma \kappa_\pm}\frac{e^{\kappa_\pm L_x/2}}{\sinh{\kappa_\pm L_x}},\\
	A_{2,\pm} &=\frac{\mathcal{F}_0}{4\gamma \kappa_\pm}\frac{e^{-\kappa_\pm L_x/2}}{\sinh{\kappa_\pm L_x}},\\
	B_{1,\pm} &=\frac{\kappa_\alpha \mathcal{F}_0}{8\gamma\kappa_\pm}\frac{1}{\sinh{\kappa_\pm L_x}}\left(\frac{1}{\kappa_\mp\sinh{\kappa_\mp L_x}}(e^{-\kappa_\pm L_x/2}-e^{\kappa_\pm L_x/2}\cosh{\kappa_\mp L_x})-\frac{L_x}{2} e^{\kappa_\pm L_x/2}\right),\\
	B_{2,\pm} &=\frac{\kappa_\alpha \mathcal{F}_0}{8\gamma\kappa_\pm}\frac{1}{\sinh{\kappa_\pm L_x}}\left(\frac{1}{\kappa_\mp\sinh{\kappa_\mp L_x}}(-e^{\kappa_\pm L_x/2}+e^{-\kappa_\pm L_x/2}\cosh{\kappa_\mp L_x})+\frac{L_x}{2} e^{-\kappa_\pm L_x/2}\right)\; .
\end{align}
\end{subequations}

%In a metal with Rashba SOC, the Josephson current is given by 
Having found the condensate function, we proceed to calculate the Josephson current in the F wire, which is given as
\begin{equation}\label{I_x_Rashba}
	I_{\mathrm{S}}=\frac{\pi \sigma_F T}{e} \sum_\omega \mathrm{Im}\left(f^*_0(\partial_x f_0-\kappa_\alpha f_2)-f^*_2\partial_x f_2\right)\; .
\end{equation}
%where $\sigma_F=e^2 N_0 D$ is the conductivity of the ferromagnet and $N_0$ is the density of states at the Fermi level.
Note that in Eq.~\eqref{I_x_Rashba} we introduce  the Matsubara  frequencies $\omega=2\pi(n+1/2)T$. The Matsubara GF is obtained by analytic continuation of $\check{f}$ to the complex plane $\epsilon \rightarrow i\omega$.
We can use the boundary conditions~\eqref{BC_Rashba_components} to simplify the previous equation:
\begin{equation}\label{current_x}
	I_{\mathrm{S}}=\frac{2\pi T}{e R_b} \sum_{\omega > 0} \mathrm{Im}f^*_0(L_x/2)\mathcal{F}_0 e^{i\phi/2}\; .
\end{equation}
After substitution  the Eqs.~\eqref{eqGF0} and \eqref{eqGF2} in Eq.~\eqref{current_x}, we find the Josephson current: 
\begin{equation}\label{I_S_rashba}
	I_{\mathrm{S}}=\frac{2 \pi \sigma_S T}{e \gamma^2} \sum_{\omega > 0} \mathcal{F}^2_0 \bigg[\mathrm{Re}\frac{1}{\kappa_+\sinh{(\kappa_+ L_x)}}\sin{\phi}+\kappa_\alpha \mathrm{Im}\left(\frac{-L_x/2}{\kappa_+\sinh{(\kappa_+ L_x)}}+\frac{\cosh{(\kappa_+ L_x)}}{|\kappa_+|^2|\sinh{(\kappa_+ L_x)}|^2}\right)\cos{\phi}\bigg].
\end{equation}
Here we  identify the usual $I_0$ and anomalous $I_\mathrm{an}$ components of the Josephson current as the terms proportional to $\sin \phi$ and $\cos \phi$, respectively. The usual component is independent of the SOC strength, while the anomalous current is proportional to $\kappa_\alpha$ and  is non-vanishing if the exchange field $h$ is finite. 

%Studying the amplitude of the anomalous Joshepson current we see that we need both a finite exchange field and Rashba SOC strength to obtain the $\phi_0$ effect.

\subsection{Current along the \texorpdfstring{$y$}{} wire \label{subsec1b}}
%The condensate function on the $y$ wire induced by the proximity effect is described by the Kuprianov-Lukichev boundary condition~\cite{Kuprianov-Lukichev}
%\begin{equation}
%	\sigma_N R_B \check{g}_y \partial_z \check{g}_y=\frac{1}{2}[\check{g}_x(0),\check{g}_y]\; ,
%\end{equation}
%where $R_B$ is the resistance per unit area between the wires. Linearizing the boundary condition:
%\begin{equation}
%	\partial_z \check{f}_y|_{z=0}=-\frac{w_y}{\gamma_0^2}\check{f}_x(0)\; ,
%\end{equation}
%where $\gamma_0^2=R_B \sigma_N w_y$. Here we have assumed that the interface resistance is much larger than the resistance of the wires, such that the condensate in the $y$ wire is determined by a small leakage from the $x$ wire. If the widths of the wires $w_{x,y}$ are much smaller than the superconducting coherence length, we can integrate the Usadel equation along the $z$ axis. The boundary condition is transferred to the equation as described
%Main text cross-reference
%in Eq.~(5) in the main text.

%The anomalous GF should vanish at the normal electrodes: $\check{f}(\pm L_y/2)=0$. Solving the boundary condition problem, we obtain the the condensate function in the $y$ wire is given by
Starting from Eq.~\eqref{Usadely}, and imposing $\check{f}_y(\pm L_y/2)=0$, we find the condensate function in the $y$-wire
\begin{equation}\label{solution2}
	\check{f}_y=\frac{w_x}{2\gamma_0^2 \sqrt{-i}\kappa'_\epsilon \cosh{(\sqrt{-i}\kappa'_\epsilon L_y}/2)}\check{f}(0)\sinh{(\sqrt{-i}\kappa'_\epsilon (L_y/2-\abs{y}))}\; ,
\end{equation}
where $\check{f}(0)$ is the condensate function for the $x$-wire (found in Sec.~\ref{subsec1A}), evaluated at the intersection of $x$ and $y$ wires.
%The perpendicular wire consists on a N-N-N structure, where the terminals are subjected to a voltage difference. If the interfacial resistance between the $x$ and $y$ wires $R_B$ is much larger than the resistance of the wires, $R_B \gg L_{x,y}/\sigma_{x,y}$, then the leakage of Cooper pairs from the $x$ wire into the $y$ wire is small and we can neglect the leakage of the current into the $x$ wire. Then, the only contribution to the current comes from the quasiparticle current, which is given by
%Main text cross-reference
%in Eq.~(6) in the main text. The corrections to the usual Ohmic current can be decomposed into:
Next, we use $\check{f}_y$ to calculate the quasiparticle current $\delta I_{\mathrm{qp}}$ from Eq.~\eqref{quasiparticle_current}. We decompose this current as $\delta I_{\mathrm{qp}}=\delta I_1+\delta I_2$, where
\begin{align}\label{I_1}
	\delta I_1&=-\frac{\sigma_N}{eL_y}\int d\epsilon F_T(\epsilon,V/2)\frac{1}{16}\left\langle\mathrm{Tr}((\check{f}^R)^2+(\check{f}^A)^2)\right\rangle=\frac{\pi T\sigma_N}{2eL_y}\mathrm{Im}\sum_{\omega>0} \langle\mathrm{Tr}\check{f}(\omega+ieV/2)^2\rangle,\\
    \label{I_2}
	\delta I_2 &=\frac{4\sigma_N}{eL_y}\int_{0}^{\infty} d\epsilon F_T(\epsilon,V/2)\frac{1}{16}\left\langle\mathrm{Tr}(\check{f}^R\check{f}^A)\right\rangle\; .
\end{align}
Using solution~\eqref{solution2}, the summands in Eq.~\eqref{I_1} can be written as
\begin{equation}\label{mw}
	\langle\mathrm{Tr}\check{f}(\omega+ieV/2)^2\rangle=\left.\frac{w_x^2}{\gamma_0^4\left.\kappa'_\omega\right.^2 \cosh^2{(\kappa'_\omega L_y}/2)}\left(\frac{\sinh{(\kappa'_\omega L_y)}}{2\kappa'_\omega L_y}-\frac{1}{2}\right)(|f_0(0)|^2-|\boldsymbol{f}(0)|^2)\right|_{\omega=2\pi(n+1/2)T+ieV/2}\; ,
\end{equation}
where $f_0(0)$ and $\boldsymbol{f}(0)$ are the singlet and triplet component of the GF of the $x$ wire at the crossing point. 
Similarly, the integrand in Eq.~\eqref{I_2} can be written as
\begin{equation}\label{mRA_Rashba}
\begin{aligned}
	\left\langle\mathrm{Tr}(\check{f}^R\check{f}^A)\right\rangle=&
	\frac{w_x^2 \left(\frac{\sinh{(\sqrt{2}\kappa'_\epsilon L_y/2)}}{\sqrt{2}\kappa'_\epsilon L_y}-\frac{\sinh{(\sqrt{2}i\kappa'_\epsilon L_y/2)}}{\sqrt{2}i\kappa'_\epsilon L_y}\right)}{2\gamma_0^4|\kappa'_\epsilon|^2 |\cosh{(\sqrt{-i}\kappa'_\epsilon L_y}/2)|^2}(f_0(\epsilon,\boldsymbol{h},\alpha)f_0(\epsilon,-\boldsymbol{h},-\alpha)^*\\
	+f_0(-\epsilon,\boldsymbol{h},\alpha)^* & f_0(-\epsilon,-\boldsymbol{h},-\alpha)-\boldsymbol{f}(\epsilon,\boldsymbol{h},\alpha)\cdot\boldsymbol{f}(\epsilon,-\boldsymbol{h},-\alpha)^*-\boldsymbol{f}(-\epsilon,\boldsymbol{h},\alpha)^*\cdot\boldsymbol{f}(-\epsilon,-\boldsymbol{h},-\alpha))\; .
\end{aligned}
\end{equation}

%For the regular terms we obtain similar expressions
%\begin{align}\label{mRR_Rashba}
%	\left\langle\mathrm{Tr}(\left.\check{f}^R\right.^2)\right\rangle=&
%	\frac{w_x^2 \left(\frac{\sinh{(\kappa'_\epsilon L_y)}}{2\kappa'_\epsilon L_y}-\frac{1}{2}\right)}{\gamma_0^4 \left.\kappa'_\epsilon\right.^2 \cosh^2{(\kappa'_\epsilon L_y/2)}}(f_0(-i\epsilon,\boldsymbol{h},\alpha)f_0(i\epsilon,\boldsymbol{h},\alpha)^*-\boldsymbol{f}(-i\epsilon,\boldsymbol{h},\alpha)\cdot\boldsymbol{f}(i\epsilon,\boldsymbol{h},\alpha)^*)\; ,\\
%    \label{mAA_Rashba}
%	\left\langle\mathrm{Tr}(\left.\check{f}^A\right.^2)\right\rangle=&
%	\frac{w_x^2 \left(\frac{\sinh{(\kappa'^*_\epsilon L_y)}}{2\kappa'^*_\epsilon L_y}-\frac{1}{2}\right)}{\gamma_0^4 \left.\kappa'^*_\epsilon\right.^2 \cosh^2{(\kappa'^*_\epsilon L_y/2)}}(f_0(i\epsilon,-\boldsymbol{h},-\alpha)f_0(-i\epsilon,-\boldsymbol{h},-\alpha)^*-\boldsymbol{f}(i\epsilon,-\boldsymbol{h},-\alpha)\cdot\boldsymbol{f}(-i\epsilon,-\boldsymbol{h},-\alpha)^*)\; ,
%\end{align}

%We can use the analytic expression of the GF of the $x$ wire to calculate the condensate function in the $y$ wire. According to Eq.~\eqref{solution2} it only depends on the GF at the center of the $x$ wire. The short range modes decay over the length $\kappa_F^{-1}$, so in the long wire limit the current on the $y$ wire only depends on the long-range triplet components.

%The $y$ wire is not subject to Rashba SOC, so the quasiparticle current is given by Eqs.~(\ref{I_1}-\ref{I_2}). 

Finally, the usual ($I^{\mathrm{qp}}_0$) and anomalous ($I^{\mathrm{qp}}_{\mathrm{an}}$) quasiparticle currents are
\begin{equation}\label{Iy_0_Rashba}
\begin{aligned}
	I_0^{\mathrm{qp}}&=\frac{\pi \sigma_N T}{2 e L_y}\mathrm{Im}\sum_{n \geq 0} \frac{w_x^2 \left(\frac{\sinh{(\kappa'_\omega L_y)}}{2\kappa'_\omega L_y}-\frac{1}{2}\right)}{\gamma_0^4\left.\kappa'_\omega\right.^2 \cosh^2{(\kappa'_\omega L_y}/2)}\frac{\mathcal{F}_{0}^2}{4\gamma^2}\left.\left(\frac{1}{\kappa_+^2\sinh^2{(\kappa_+ L_x/2)}}+\frac{1}{\kappa_-^2\sinh^2{(\kappa_- L_x/2)}}\right)\right|_{\omega=2\pi(n+1/2)T+ieV/2}\\
	&+\frac{\sigma_N}{4eL_y}\int_{0}^{\infty} d\epsilon F_T(\epsilon,V/2)\frac{w_x^2 \left(\frac{\sinh{(\sqrt{2}\kappa'_\epsilon L_y/2)}}{\sqrt{2}\kappa'_\epsilon L_y}-\frac{\sinh{(\sqrt{2}i\kappa'_\epsilon L_y/2)}}{\sqrt{2}i\kappa'_\epsilon L_y}\right)}{\gamma_0^4|\kappa'_\epsilon|^2 |\cosh{(\sqrt{-i}\kappa'_\epsilon L_y}/2)|^2}\frac{|\mathcal{F}_0|^2}{4\gamma^2}\\
	& \hspace{9cm} \cdot \left(\frac{1}{|\kappa_{+}|^2|\sinh{(\kappa_{+} L_x/2)}|^2}+\frac{1}{|\kappa_{-}|^2|\sinh{(\kappa_{-} L_x/2)}|^2}\right),
\end{aligned}
\end{equation}
\begin{equation}\label{Iy_an_Rashba}
\begin{aligned}
	I_{\mathrm{an}}^{\mathrm{qp}}&=\kappa_\alpha\frac{\pi \sigma_N T}{2 e L_y}\mathrm{Im}\sum_{n \geq 0} \frac{w_x^2 \left(\frac{\sinh{(\kappa'_\omega L_y)}}{2\kappa'_\omega L_y}-\frac{1}{2}\right)}{\gamma_0^4\left.\kappa'_\omega\right.^2 \cosh^2{(\kappa'_\omega L_y}/2)}\frac{\mathcal{F}_{0}^2}{4\gamma^2}  \left(\frac{L_x}{2\kappa_+^2\sinh^2{(\kappa_+ L_x/2)}}-\frac{L_x}{2\kappa_-^2\sinh^2{(\kappa_- L_x/2)}}\right.\\
	& \hspace{6cm} \left.\left.+\frac{\tanh{(\kappa_- L_x/2)}}{\kappa_+^2\kappa_-\sinh^2{(\kappa_+ L_x/2)}}-\frac{\tanh{(\kappa_+ L_x/2)}}{\kappa_-^2\kappa_+\sinh^2{(\kappa_- L_x/2)}}\right)\right|_{\omega=2\pi(n+1/2)T+ieV/2}\\
	&+\kappa_\alpha\frac{\sigma_N}{4eL_y}\int_{0}^{\infty} d\epsilon F_T(\epsilon,V/2)\frac{w_x^2 \left(\frac{\sinh{(\sqrt{2}\kappa'_\epsilon L_y/2)}}{\sqrt{2}\kappa'_\epsilon L_y}-\frac{\sinh{(\sqrt{2}i\kappa'_\epsilon L_y/2)}}{\sqrt{2}i\kappa'_\epsilon L_y}\right)}{\gamma_0^4|\kappa'_\epsilon|^2 |\cosh{(\sqrt{-i}\kappa'_\epsilon L_y}/2)|^2}\frac{|\mathcal{F}_0|^2}{4\gamma^2}\\
	& \hspace{6cm} \cdot\; \mathrm{Im}\left\{\frac{\tanh{(\kappa_- L_x/2)}}{|\kappa_+|^2\kappa_-|\sinh{(\kappa_+ L_x/2)}|^2}-\frac{\tanh{(\kappa_+ L_x/2)}}{|\kappa_-|^2\kappa_+|\sinh{(\kappa_- L_x/2)}|^2}\right\}.
\end{aligned}
\end{equation}
%Comparing the quasiparticle current to the Josephson current in the $x$ wire, we can identify $I_0^{\mathrm{qp}}$, which is independent of $\kappa_\alpha$, with the usual Josephson current and $I_{\mathrm{an}}^{\mathrm{qp}}$, which is proportional to $\kappa_\alpha$ with the anomalous Josephson current.
%Main text cross-reference
The  factors $c_1$, and $c_2$ defined in Eqs.~(\ref{I^qp_0_2}-\ref{I^qp_an_2}) of the main text can be extracted from Eqs.~(\ref{Iy_0_Rashba}-\ref{Iy_an_Rashba}) by assuming the limit of weak exchange field $h$.

\section{S/FI/F/FI/S junction \label{sec2}}
In this Appendix we present a detailed derivation of the expressions used in the main text for the currents in the S/FI/F/FI/S geometry. 
In Sec.~\ref{spin-filter GF x} we first present the derivation of the Josephson current in the $x$-wire, followed by the derivation of the quasiparticle current in the $y$-wire in Sec.~\ref{spin-filter GF y}.

\subsection{Current along the \texorpdfstring{$x$}{} wire}\label{spin-filter GF x}
%The second configuration we study to obtain a $\phi_0$-junction is a S/FI/F/FI/S junction with non-coplanar magnetizations. We can study the weak proximity effect in the F layer by linearizing the Usadel equation with respect to the anomalous Green’s function $\check{f}$. We assume, without loss of generality, that the exchange field lies in the $\boldsymbol{z}$ axis $\boldsymbol{h}=h \boldsymbol{z}$. In this case, the Usadel equation in the $x$ wire is given by
%Main text cross-reference
%Eq.~(9) in the main text.

The general solution of Eq.~\eqref{Usadel linearized} for the condensate function in the $x$-wire reads
\begin{equation}\label{solution1}
	\hat{f} =(A+A\sigma_3)\mathrm{e}^{\kappa_+ x} + (B+B\sigma_3)\mathrm{e}^{-\kappa_+ x} + (C-C\sigma_3)\mathrm{e}^{\kappa_- x} + (D-D\sigma_3)\mathrm{e}^{-\kappa_- x}+E\sigma_1 \mathrm{e}^{i \kappa_\epsilon x}+F\sigma_1 \mathrm{e}^{-i\kappa_\epsilon x}+G\sigma_2 \mathrm{e}^{i\kappa_\epsilon x}+H\sigma_2 \mathrm{e}^{-i\kappa_\epsilon x}\; .
\end{equation}
Coefficients in Eq.~\eqref{solution1} can be found by applying the boundary conditions [Eq.~\eqref{boundary_condition_linearized}].
%The spin-filtering barriers are described by the generalized
%Kuprianov-Lukichev boundary conditions~\cite{Bergeret:2012,Eschrig:2015}. The linearized boundary condition is greatly simplified when the exchange field on the superconducting electrodes is collinear with the barrier polarization $\boldsymbol{n}_{r/l}=\boldsymbol{m}_{r/l}$~\cite{Silaev:2017}
%Main text cross-reference
%(see Eq.~(10) in the main text).
%No voltage is applied on the SF-F-FS wire, so the current is given by the condensate current, which can be written in terms of $\check{f}_\omega$
%\begin{equation}\label{condensate_current}
%	I_{\mathrm{S}}=\frac{\sigma_F}{16e}\int d\epsilon\, \mathrm{Tr}\{\tau_3(\check{g}^{R}\partial_x\check{g}^{R}-\check{g}^{A}\partial_x\check{g}^{A})F_L\}=i\pi\sigma_F\frac{T}{4e}\sum_{\omega} \mathrm{Tr}\{\tau_3\check{g}_\omega \partial_x \check{g}_\omega\}\; .
%\end{equation}
%In this case, the distribution function $F_L=\tanh{\epsilon/(2T)}$ is diagonal in Nambu space.
In the ferromagnetic wire, the supercurrent is given as
\begin{equation}\label{current1}
	I_{\mathrm{S}}=\pi\sigma_F\frac{T}{e}\sum_{\omega} \mathrm{Im}\{f_0^* \partial_x f_0-\boldsymbol{f}^* \cdot\partial_x \boldsymbol{f}\}\; ,
\end{equation}
where $\hat{f}_\omega=f_0+\boldsymbol{f}\cdot\boldsymbol{\sigma}$ is decomposed into the scalar singlet amplitude $f_0$ and the vector of triplet states $\boldsymbol{f}$.
Assuming $\kappa_F^{-1} \ll L_x \ll \kappa_\omega^{-1}$, we can substitute the long range components $f_1$ and $f_2$ by their average values, given by
\begin{equation}
	\langle f_{1/2} \rangle=\frac{\partial_x f_{1/2}|_{x=-L_x/2}-\partial_x f_{1/2}|_{x=L_x/2}}{\kappa_\omega^2 L_x}\; .
\end{equation}
Then, using the boundary conditions [Eq.~\eqref{boundary_condition_linearized}], the long range triplet components are given by
\begin{equation}\label{f1_2}
\begin{aligned}
	\langle f_{1/2} \rangle=&\frac{1}{\kappa_\omega^2 L_x \gamma}(i\mathcal{G}_{l,0}(P_{l,2/3}f_{3/1}(-L_x/2)-P_{l,3/1}f_{2/3}(-L_x/2))-\sqrt{1-P_l^2}\mathcal{F}_{l,1/2}e^{-i \phi/2}\\
&+i\mathcal{G}_{r,0}(P_{r,2/3}f_{3/1}(L_x/2)-P_{r,3/1}f_{2/3}(L_x/2))-\sqrt{1-P_r^2}\mathcal{F}_{r,1/2}e^{i \phi/2})\; ,
\end{aligned}
\end{equation}
where $\hat{\mathcal{G}}_{r/l}$ and $\hat{\mathcal{F}}_{r/l}$ are given by Eq.~\eqref{GF S electrodes}.
%$\hat{\mathcal{G}}_{r/l} \equiv \mathcal{G}_{r/l,0}+\boldsymbol{\mathcal{G}_{r/l}}\cdot\boldsymbol{\sigma}$ and $\hat{\mathcal{F}}_{r/l} \equiv \mathcal{F}_{r/l,0}+\boldsymbol{\mathcal{F}_{r/l}}\cdot\boldsymbol{\sigma}$ are the normal and anomalous GF of the superconducting electrodes. If the exchange field in the electrodes is small, their GF is given by Eq.~(11) in the main text.
%Main text cross-reference
Using solution~\eqref{solution1}, we can calculate the $f_0$ and $f_3$ near each boundary independently without overlapping. To first order in $\gamma^{-1}$, we obtain:
\begin{align}\label{f0_1}
	f_{0,r/l}^{(1)}&=\frac{\sqrt{1-P_{r/l}^2}}{2\gamma}\left(\frac{\mathcal{F}_{r/l,0}+\mathcal{F}_{r/l,3}}{\kappa_+}e^{-\kappa_+(L_x/2 \mp x)}+\frac{\mathcal{F}_{r/l,0}-\mathcal{F}_{r/l,3}}{\kappa_-}e^{-\kappa_-(L_x/2 \mp x)}\right)e^{i \eta_{r/l}\phi/2}\\
	\label{f3_1}
	f_{3,r/l}^{(1)}&=\frac{\sqrt{1-P_{r/l}^2}}{2\gamma}\left(\frac{\mathcal{F}_{r/l,0}+\mathcal{F}_{r/l,3}}{\kappa_+}e^{-\kappa_+(L_x/2 \mp x)}-\frac{\mathcal{F}_{r/l,0}-\mathcal{F}_{r/l,3}}{\kappa_-}e^{-\kappa_-(L_x/2 \mp x)}\right)e^{i \eta_{r/l}\phi/2}\; .
\end{align}
The second-order terms in $\gamma^{-1}$ are also important to obtain the anomalous Josephson effect:
\begin{align}\label{f0_2}
	f_{0,r/l}^{(2)}&=\frac{-i \mathcal{G}_{r/l,0}}{2\gamma}(P_{r/l,1}\langle f_2 \rangle-P_{r/l,2}\langle f_1 \rangle)\left(\frac{e^{-\kappa_+(L_x/2 \mp x)}}{\kappa_+}-\frac{e^{-\kappa_-(L_x/2 \mp x)}}{\kappa_-}\right)\\
	\label{f3_2}
	f_{3,r/l}^{(2)}&=\frac{-i \mathcal{G}_{r/l,0}}{2\gamma}(P_{r/l,1}\langle f_2 \rangle-P_{r/l,2}\langle f_1 \rangle)\left(\frac{e^{-\kappa_+(L_x/2 \mp x)}}{\kappa_+}+\frac{e^{-\kappa_-(L_x/2 \mp x)}}{\kappa_-}\right)\; .
\end{align}
To lowest order in $\gamma^{-1}$, the current~\eqref{current1} can be written as
\begin{equation}\label{current2}
	I_{\mathrm{S}}=\frac{\pi\sigma_F T}{e\gamma}\sum_{\omega} \mathrm{Im}\pm\sqrt{1-P_{r/l}^2}(f_0^* \mathcal{F}_{r/l,0}-\boldsymbol{f}^* \cdot\boldsymbol{\mathcal{F}}_{r/l})e^{i \eta_{r/l}\phi/2}|_{x=\pm L_x/2}\; .
\end{equation}

%An analytic expression for the GF can be obtained in the $\kappa_F^{-1} \ll L_x \ll \kappa_\omega^{-1}$ regime. The short range modes decay over the length $\kappa_F^{-1}$, so in this regime the Josephson current is mediated by the long-range triplet superconducting correlations. The current can be decomposed as
%\begin{equation}
	%I_{\mathrm{S}}=I^{\mathrm{S}}_0\sin\phi+I^{\mathrm{S}}_{\mathrm{an}}\cos\phi
%\end{equation}
%$I^{\mathrm{S}}_0$ is the usual Josephson current and $I^{\mathrm{S}}_{\mathrm{an}}$ is the anomalous current. The components are given by
Finally, we find the usual and anomalous Josephson currents:
\begin{align}\label{usual_current_1}
	I^{\mathrm{S}}_0 &=-\frac{2\pi}{eR_b}(\boldsymbol{h}_{l\perp}\cdot\boldsymbol{h}_{r\perp})\frac{\sqrt{1-P_r^2}\sqrt{1-P_l^2}}{\sqrt{2}L_x\gamma^2}\sum_{\omega>0} \frac{T \mathcal{F}'^{2}_{0}}{\kappa_N^2}\\
	\label{anomalous_current_1}
	I^{\mathrm{S}}_{\mathrm{an}} &=-\frac{2\pi}{eR_b}(\chi_l-\chi_r)\frac{\sqrt{1-P_r^2}\sqrt{1-P_l^2}}{\kappa_F hL_x\gamma^3}\sum_{\omega>0} \frac{T \mathcal{G}_0 \mathcal{F}_0 \mathcal{F}'_{0}}{\kappa_\omega^2}\; .
\end{align}
For simplicity  we have assumed  the same amplitude of the order parameter in the two electrodes. The chiralities are defined as $\chi_{r/l}=\boldsymbol{h}\cdot(\boldsymbol{P_{r/l}}\cross\boldsymbol{h_{l/r}})$, and $\boldsymbol{h}_{r/l\perp}=\boldsymbol{h}_{r/l}-(\boldsymbol{h}_{r/l}\cdot\boldsymbol{h})\boldsymbol{h}/h^2$ are the components of $\boldsymbol{h}_{r/l}$ perpendicular to $\boldsymbol{h}$.
%and $\kappa_N=\sqrt{T/D}$. 
%The anomalous current smaller than $I_0$ by a factor of $(\kappa_F\gamma)^{-1}$, so the total current will dominantly be given by $I_0$ unless $\boldsymbol{h}_{l\perp}\cdot\boldsymbol{h}_{r\perp}=0$. In this case, the anomalous current dominates and the current is shifted by a phase $\phi_0=\pi/2$.

\subsection{Current along the \texorpdfstring{$y$}{} wire \label{spin-filter GF y}}

%We obtain that the regular part of the current has a constant component and a part dependent on the phase $\phi$ (the anomalous part has a similar phase structure). The most interesting part is the one depend on the phase, which can be decomposed into two components:
We obtain the quasiparticle current in the $y$-wire by following the same procedure as in Sec.~\ref{subsec1b}. The coefficients $c_3$ and $c_4$ in Eqs.~\eqref{I^qp_0_1} and~\eqref{I^qp_an_1} are:
%\begin{equation}\label{Iy_reg}
%	\delta I_{\mathrm{qp}}=\sqrt{1-P_l^2}\sqrt{1-P_r^2}(c_3 \gamma^{-2}\boldsymbol{h}_{l\perp}\cdot\boldsymbol{h}_{r\perp}\cos{\phi}+c_4\gamma^{-3}(\chi_l-\chi_r)\sin{\phi})\; ,
%\end{equation}
\begin{align}\label{c_3}
	c_3&=-\frac{\pi T\sigma_N}{2eL_y}\mathrm{Im}\sum_{n \geq 0} \frac{w_x^2 \left(\frac{\sinh{(\kappa'_\omega L_y)}}{2\kappa'_\omega L_y}-\frac{1}{2}\right)}{\gamma_0^4\left.\kappa'_\omega\right.^2 \cosh^2{(\kappa'_\omega L_y}/2)} \left.\frac{2(\mathcal{F}'_{0})^2}{\kappa_\omega^4 L_x^2}\right|_{\omega=2\pi(n+1/2)T+ieV/2}\nonumber \\
	&+\frac{\sigma_N}{4eL_y}\int_{0}^{\infty} d\epsilon F_T(\epsilon,V/2)\frac{w_x^2 \left(\frac{\sinh{(\sqrt{2}\kappa'_\epsilon L_y/2)}}{\sqrt{2}\kappa'_\epsilon L_y}-\frac{\sinh{(\sqrt{2}i\kappa'_\epsilon L_y/2)}}{\sqrt{2}i\kappa'_\epsilon L_y}\right)}{\gamma_0^4|\kappa'_\epsilon|^2 |\cosh{(\sqrt{-i}\kappa'_\epsilon L_y}/2)|^2}\frac{2|\mathcal{F}'_0|^2}{|\kappa_\epsilon|^4 L_x^2}\; ,
\end{align}
\begin{align}\label{c_4}
	c_4&=\frac{\pi T\sigma_N}{2eL_y}\mathrm{Im}\sum_{n \geq 0} \frac{w_x^2 \left(\frac{\sinh{(\kappa'_\omega L_y)}}{2\kappa'_\omega L_y}-\frac{1}{2}\right)}{\gamma_0^4\left.\kappa'_\omega\right.^2 \cosh^2{(\kappa'_\omega L_y}/2)}\left.\frac{\sqrt{2}\mathcal{G}_0\mathcal{F}_{0}\mathcal{F}'_{0}}{\kappa_F \kappa_\omega^4 L_x^2}\right|_{\omega=2\pi(n+1/2)T+ieV/2} \nonumber \\
	&+\frac{\sigma_N}{4eL_y}\int_{0}^{\infty} d\epsilon F_T(\epsilon,V/2)\frac{w_x^2 \left(\frac{\sinh{(\sqrt{2}\kappa'_\epsilon L_y/2)}}{\sqrt{2}\kappa'_\epsilon L_y}-\frac{\sinh{(\sqrt{2}i\kappa'_\epsilon L_y/2)}}{\sqrt{2}i\kappa'_\epsilon L_y}\right)}{\gamma_0^4|\kappa'_\epsilon|^2 |\cosh{(\sqrt{-i}\kappa'_\epsilon L_y}/2)|^2}\frac{-\sqrt{2}\mathrm{Re}\{\mathcal{G}^*_0\mathcal{F}^*_0\mathcal{F}'_0\}}{\kappa_F |\kappa_\epsilon|^4 L_x^2}\; .
\end{align}

\twocolumngrid

%\nocite{*}
%\bibliography{biblio}\label{LastBibItem}


\end{document}